\documentclass[twocolumn]{aastex62}
\usepackage{amsmath}
\usepackage{graphicx}
\usepackage{color}
\usepackage{amsmath}
\usepackage{graphicx}

\newcommand{\beq}{\begin{equation}}
\newcommand{\eeq}{\end{equation}}

\begin{document}

\title{Evolution of the Black Hole Mass Function in Star Clusters from Multiple Mergers}

\correspondingauthor{Pierre Christian}
\email{pchristian@cfa.harvard.edu}

\author{Pierre Christian}
\affiliation{Astronomy Department, Harvard University, 60 Garden St., Cambridge, MA 02138}

\author{Philip Mocz}
\affiliation{Department of Astrophysical Sciences, Princeton University, 4 Ivy Lane, Princeton, NJ, 08544, USA}
\altaffiliation{Einstein Fellow}

\author{Abraham Loeb}
\affiliation{Astronomy Department, Harvard University, 60 Garden St., Cambridge, MA 02138}

%\documentclass{emulateapj}
%\usepackage{amsmath}
%\usepackage{graphicx}
%\usepackage{color}
%\usepackage{amsmath}
%\usepackage{graphicx}
%\usepackage[breaklinks,colorlinks,citecolor=blue]{hyperref}

%\begin{document}

%\title{Evolution of the Black Hole Mass Function in Star Clusters from Multiple Mergers}
%\author{Pierre Christian$^1$, Philip Mocz$^{2\dagger}$, and Abraham Loeb$^1$}
%\affiliation{$^1$Astronomy Department, Harvard University, 60 Garden St., Cambridge, MA 02138\\
%$^2$Department of Astrophysical Sciences, Princeton University, 4 Ivy Lane, Princeton, NJ, 08544, USA}
%\email{pchristian@cfa.harvard.edu; aloeb@cfa.harvard.edu}
%\thanks{$^\dagger$ Einstein Fellow}

\begin{abstract} %Note:abstract changed
We investigate the effects of black hole mergers in star clusters on the black hole mass function. As black holes are not produced in pair-instability supernovae, it is suggested that there is a dearth of high mass stellar black holes. This dearth generates a gap in the upper end of the black hole mass function. Meanwhile, parameter fitting of X-ray binaries suggests the existence of a gap in the mass function under $5$ solar masses. We show, through evolving a coagulation equation, that black hole mergers can appreciably fill the upper mass gap, and that the lower mass gap generates potentially observable features at larger mass scales. We also explore the importance of ejections in such systems and whether dynamical clusters can be formation sites of intermediate mass black hole seeds.\\
\end{abstract}
%\maketitle

\section{Introduction}
The discovery of merging black holes (BHs) by the Laser Interferometer Gravitation-Wave Observatory (LIGO) signaled the beginning of gravitational wave astrophysics \citep{LIGO1,2016PhRvL.116x1103A, 2017PhRvL.118v1101A,2017PhRvL.119n1101A,2017ApJ...851L..35A}. The masses of these binaries are much larger than those previously discovered as X-ray binaries \citep{2010ApJ...725.1918O,2011ApJ...741..103F,2012ApJ...757...36K}. The existence of these massive BHs was anticipated by previous calculations of BH mergers \citep{2010ApJ...714.1217B,2010ApJ...715L.138B,2015ApJ...806..263D}, and their detection spurred a growing interest on their formation mechanisms. One promising mechanism that allows binary BHs of such masses to form is the dynamical merger scenario, where BHs in dense star clusters gravitationally interact with each other to produce very hard binaries \citep{2006ApJ...637..937O,2015PhRvL.115e1101R,2016PhRvD..93h4029R,2016ApJ...824L...8R,2016ApJ...821...38S,2018arXiv180208654S}. 

In such systems, BHs borne out of mergers can merge again producing second generation BHs \citep{2016ApJ...831..187A,2017PhRvD..95l4046G,2017ApJ...840L..24F,2017arXiv171204937R}. These multiple mergers necessarily increase the number of massive BHs while simultaneously lowering the number of less massive BHs, turning the BH mass function (BHMF) more top-heavy in the process.  

%While the lower bound of his mass gap could be higher for BHs in dense clusters,
Supernova theory also predicts the existence of a mass gap in the BH initial mass function (IMF) between $50-130 M_\odot$ because the stellar progenitors of BHs in this mass range undergo pair-instability supernovae \citep{2016A&A...594A..97B, 2017ApJ...836..244W}. Recently, parameter fitting of four LIGO data points suggests the existence of a cutoff at $M \sim 40 M_\odot$, bolstering the validity of this theoretical calculation \citep{2017ApJ...851L..25F}. Further, more massive binary BHs can be observed by LIGO to a greater distance, and so the absence of LIGO events at $M \gtrapprox 40 M_\odot$ within the increased survey volume can be used to set an upper limit on the BHMF. Analysis on the redshift distribution of LIGO events corroborates the existence of this mass gap \citep{2018arXiv180204909B}.

In the dynamical merger scenario, multiple merger events might be able to appreciably fill the upper mass gap. In addition, while binary BHs in isolated binaries can merge to produce BHs in the upper mass gap, the lack of multiple merger events results in a very different BHMF within the upper mass gap.  As such, the BHMF within the upper mass gap could be an effective test for the dynamical merger scenario.

%Sukhbold, private correspondence states: "The main reasons why the theory produces the gap between NS and BH distributions 2<Mrem<4 Msun, is twofold: (1) the neutrino-driven mechanism works like a 'switch'. It either successfully blows up to leave a NS, or it implodes to form a BH. It only very rarely experiences a significant fallback, which can produce a small BH in a successful explosion. (2) currently the smallest mass progenitor that is predicted to potentially implode is about Mzams ~ 15 Msun. All smaller stars never implode, they all consistently explode. In stars with Mzams ~ 15 Msun, the He-core is about ~4Msun (the mass that will end up in BH if it implodes)."
Finally, parameter fitting of X-ray binaries suggests the existence of a lower mass gap in the BHMF between the most massive neutron stars and the least massive BHs \citep{2010ApJ...725.1918O,2011ApJ...741..103F}. Under certain scenarios, supernova explosions can naturally produce this gap \citep{2012ApJ...757...91B}. According to Ref. \cite{2012ApJ...757...91B}, Rayleigh-Taylor instabilities could appear early after the initial bounce of a supernova, and drive explosions $\lesssim100-200$ ms after the collapse. In such rapid explosions, stars of mass $\sim 14-20 M_\odot$ are thought to produce strong explosions that result in high mass neutron stars ($M \sim 1.5-2 M_\odot$). However, stars of mass $\sim 20-40 M_\odot$ fail to explode in this scenario, forming BHs of mass $M\sim 5-10 M_\odot$.

This gap has also been successfully reproduced by numerical simulations of neutrino-driven supernova explosions \citep{2016ApJ...821...38S}. Neutrino-driven explosions suggests that smaller stars never implode to form BHs, a prediction that has also been corroborated by the observed BH and neutron star distributions \citep{2017arXiv171200021R}. If multiple mergers are allowed, the lack of BHs in this range will have repercussions to the BHMF even at larger mass scales, as heavier BHs cannot merge with BHs in the lower mass gap to produce more massive BHs.

To answer such questions, a method to quickly compute the evolution of the BHMF is needed. In this work, we will employ the Smoluchowski coagulation equation \cite{Smoluchowsky}, a rate equation describing the time evolution of the number of particles of a certain size as the particles are allowed to interact and `coagulate', merging to form larger particles. The calculations performed in this formalism are much faster than those required in N-body simulations of dynamical clusters, allowing a large parameter space to be covered efficiently.    

This paper is organized as follows: in Section 2 we describe the Smoluchowski coagulation equation formalism, in Section 3 we discuss our results for the evolution of the BHMF assuming constant kernel. Subsequently Section 4 presents our results with top-heavy kernels. Finally Section 5 summarizes our conclusions.

\section{Methods}
%Note: you can get nan if dt = t_end/N and N is too small! Note: N~Mass Function, which is a terrible name of the variable, since the number or BHs (the thing usually denoted "N") is in our definition NdM.
 The evolution of the BH mass function due to mergers can be modeled by a coagulation equation,
\begin{align} \label{eq:Smol}
&\frac{\partial N(M,t)}{\partial t} = \nonumber \\
&\;\;\;\;\;\; \frac{1}{2} \int_0^M K(M-M',M') N(M-M',t)N(M',t) {\rm{d}}M' \nonumber \\
&\;\;\;\;\;\; - \int_0^\infty K(M,M') N(M,t) N(M',t) {\rm{d}}M' - S(M,t)\; ,
\end{align}
where $N(M,t) \rm{d} M$ is the number of black holes in the star cluster of mass $\in [M,M+\rm{d} M)$ at time $t$, $K(x,y)$ -- the \emph{coagulation kernel} -- is the rate of two BHs of masses $x$ and $y$ to merge, and $S(M,t)$ represents possible source or sink terms. 
The first term of this equation describes 
BHs of mass $<M$ merging to form BHs of mass $M$, while the second term describes removal of
BHs of mass $M$ merging to form BHs with mass $>M$.

Equation (\ref{eq:Smol}) is called the \cite{Smoluchowsky} coagulation equation, a general integro-differential equation that describes the statistical time-evolution of the distribution (as a function of mass, size, etc.) of a coagulating population of objects. 
The detailed physics of the coagulation process is encoded in the coagulation kernel, allowing one to just evolve the statistical ensemble.
The numerical method used to solve Equation (\ref{eq:Smol}) is described in Appendix~\ref{appendix1}.

We solve the coagulation equation for a variety of physical scenarios, and study its evolution over $10$ Gyrs. Our IMF follows the Salpeter function ($N\propto M^{\alpha}$, $\alpha=-2.35$; \citealt{1955ApJ...121..161S}) with an upper mass gap for $50 M_\odot < M < 130 M_\odot$ and a lower mass gap for $M < 5 M_\odot$. The different scenarios considered in this paper are summarized in Table 1.

\begin{table*}\label{sumpara}
\begin{center} 
\begin{tabular}{ || c|c |c|c|c |c  || } 
 \hline
   $N_{BH}$ & $R_{\rm{tot}}$ [$100$ Gpc$^{-3}$ yr$^{-1}$ ] & $N_C$ & $f_{\rm ej}$ & Kernel & Figure Number \\ 
 \hline
 1000      &         $100$         &      100    &        0         &              Constant        &  Figure \ref{fig:constantK} (top)  \\ 
 1000     &         $300$         &      33      &         0         &              Constant        &  Figure \ref{fig:constantK} (bottom)  \\ 
 100      &         $100$         &      100      &        0         &              Constant        &  Figure \ref{fig:fewBH}  \\ 
  1000      &         $300$         &      33      &        0.5         &              Constant        &  Figure \ref{fig:eject} (top)  \\ 
  1000      &         $300$         &      33      &        0.9        &              Constant        &  Figure \ref{fig:eject} (middle)  \\ 
  1000      &         $300$         &      1      &        0.9        &              Constant        &  Figure \ref{fig:eject} (bottom)  \\ 
  1000      &         $100$         &      100      &        0.9        &              Equation (\ref{gravrad})        &  Figure \ref{fig:topheavy}  \\ 
    1000      &         $10$         &      100      &        0.5        &              Equation (\ref{eq:3body})             &  Figure \ref{fig:3body}  \\ 
 \hline
\end{tabular}
\end{center} 
\caption{List of parameters for the different scenarios under consideration in this paper. $N_{BH}$ is the number of BHs per cluster, $R_{\rm{tot}}$ is the cosmological LIGO rate, $N_C$ is the number of clusters per galaxy, $f_{\rm ej}$ is the ejection fraction, and Kernel denotes the coagulation kernel we used. }
\end{table*}

\section{Constant kernel evolution}
First, we study the evolution of the BHMF assuming that the kernel $K(M,M')=K$ is a constant. This assumption is equivalent to the statement that the merger probability of two BHs is independent of their masses. In order to calibrate the constant $K$, we enforce the condition that the merger rate is equal to the LIGO merger rate per cluster, $R_{\rm{LIGO}}$. This is done by noting that the total number of mergers per unit time is
\begin{align}
R_{\rm LIGO} &= \int_0^\infty \int_0^\infty K(x,y) n(x,t) n(y,t) {\rm{d}}x {\rm{d}}y \nonumber \\
& = K\times N_{\rm BH}^2 \; , \label{eq:RKrelationship}
\end{align}
where $N_{\rm BH}$ is the number of BHs in the cluster. To estimate the LIGO rate per cluster, we adopt $\sim 10^{-2}$ per comoving Mpc$^{3}$  as the number density of Milky Way-like galaxies \citep{2009MNRAS.399.1106M}. 
% $100$ Gpc$^{-3}$ yr$^{-1}$ 
% the Galactic merger rate by adopting the number of Milky Way-like galaxies to be $10^{-2}$ per comoving Mpc$^{3}$ \citep{MWEG2}.
Adopting $100$ Gpc$^{-3}$ yr$^{-1}$ as a fiducial LIGO inferred merger rate gives the LIGO Galactic merger rate to be $N_{\rm{MWEG}} \sim 10^{-5} R_{100}$ mergers per galaxy per year. For $R_{\rm tot}$ being the reported LIGO rate, the LIGO rate per cluster is therefore given by
 \begin{align}
 R_{\rm{LIGO}} &= \frac{R_{\rm{tot}}}{N_{C} N_{\rm{MWEG}}}  \nonumber \\
  &= 10^{-5} \left[ \frac{R_{\rm{tot}}}{ 100 \; \rm{Gpc}^{-3}\; \rm{yr}^{-1}} \right] \left[ \frac{1}{N_C} \right] \; \rm yr^{-1} \; ,
 \end{align}
 where $N_C$ is the number of star clusters per galaxy. 
 
 \subsection{No ejections}
The simplest system we can study using this formalism is obtained by setting $S(M,t)=0$ in Equation (\ref{eq:Smol}). This is equivalent to saying that the BHs exist in a closed system, and that no mergers are violent enough to eject BHs out of the system. This situation is expected in cases where the star cluster is massive enough that the merger kick velocities are small compared to the escape velocity, e.g. for a star cluster at the core of a galaxy. Regardless of its limited usability, this simple case illustrates a lot of general features that are also present in more complicated cases. 
 
 \begin{figure}
\centering
\includegraphics[width=3.5in]{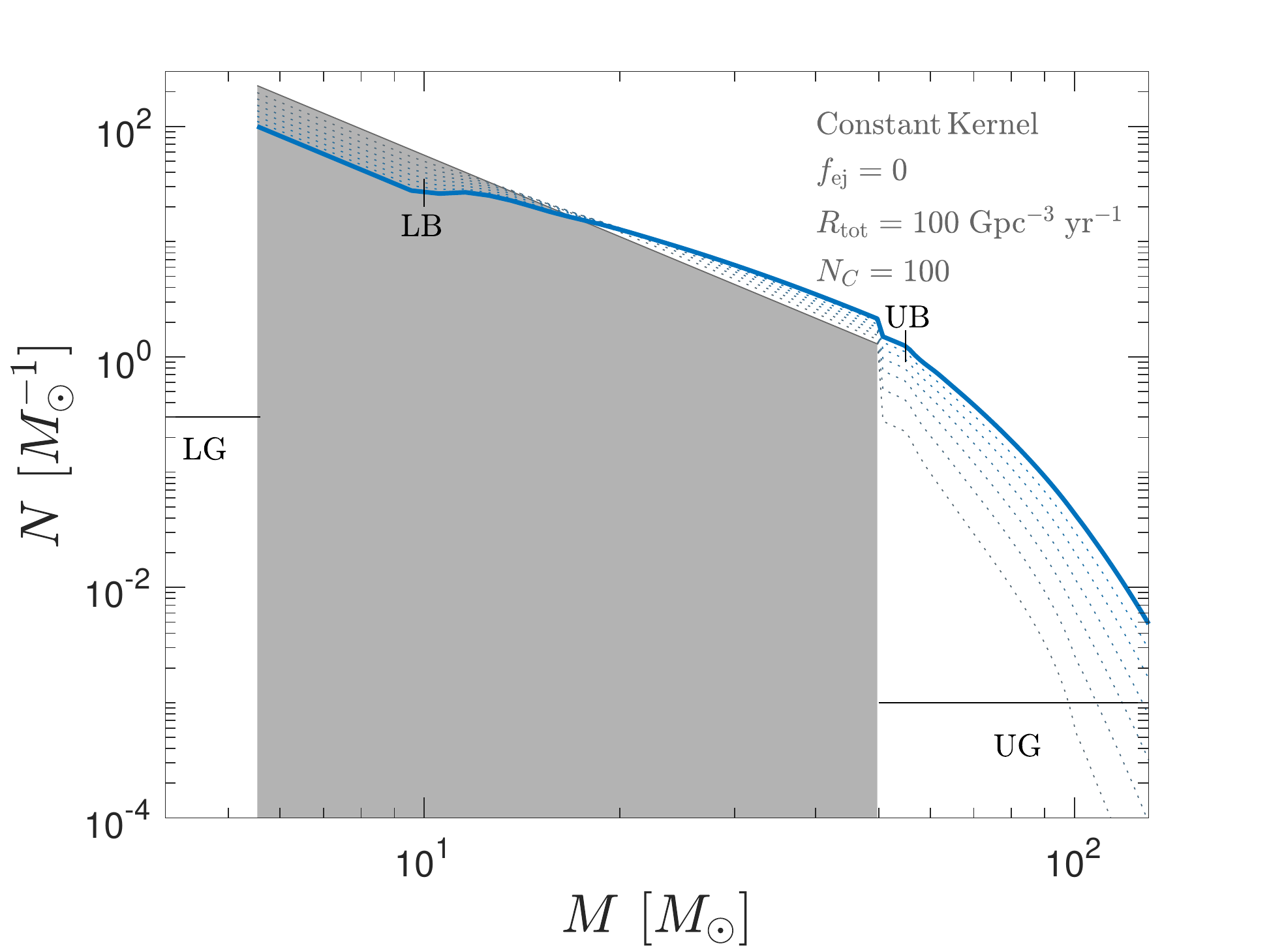} \\
\includegraphics[width=3.5in]{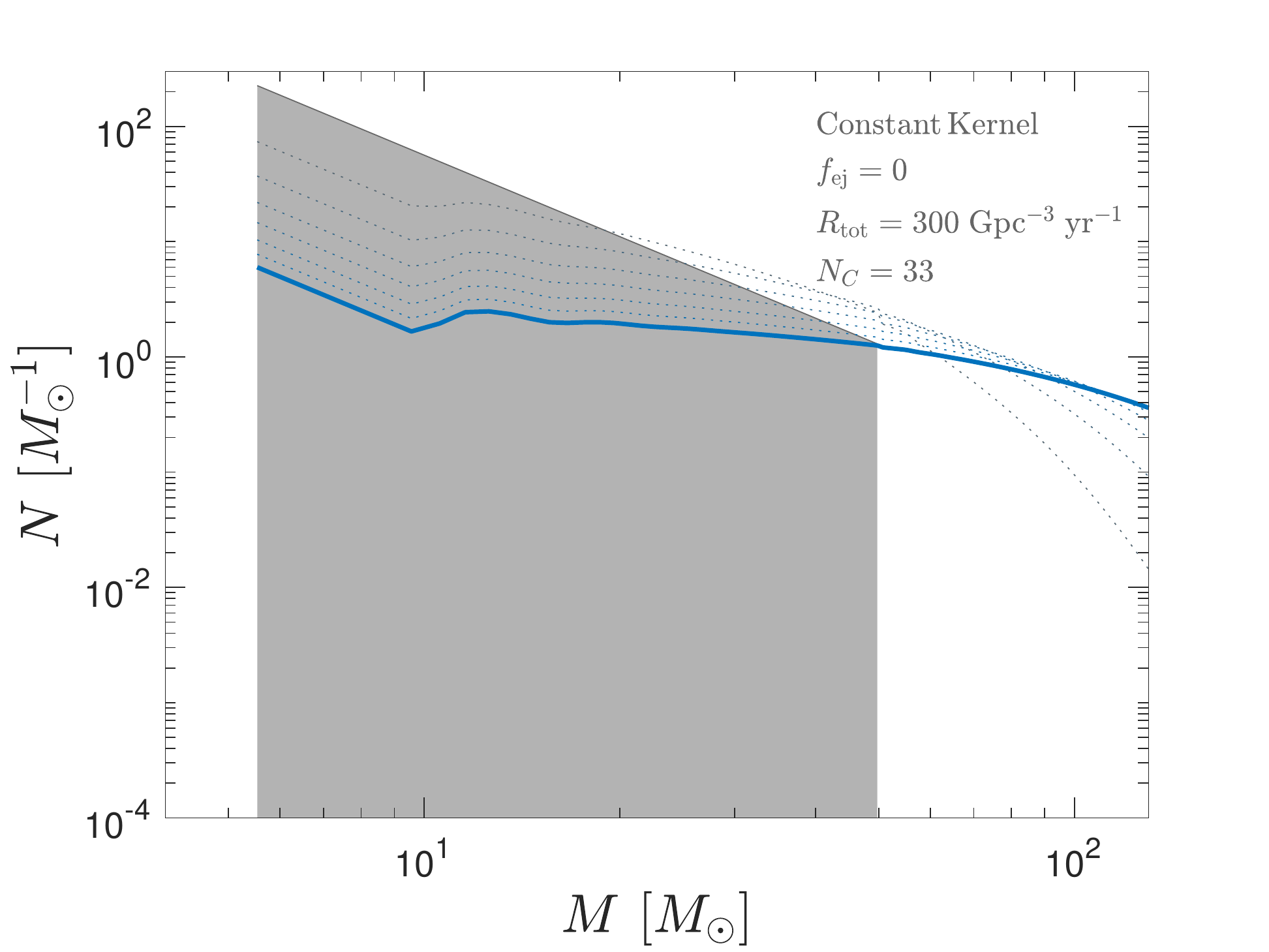}
\caption{The evolution of the BHMF starting from the IMF (black) to $10$ Gigayears (solid blue) for a cluster with $1000$ BHs. Dotted blue lines represent the BHMF at intervening times. The top figure shows evolution of the mass function with a LIGO rate of $100$ Gpc$^{-3}$ yr$^{-1}$ and the number of clusters per MWEG to be $N_C=100$, while the lower figure shows evolution of the mass function with a LIGO rate of $300$ Gpc$^{-3}$ yr$^{-1}$ and $N_C = 33$, i.e. a LIGO rate \emph{per cluster} that is $\sim$10 times higher. Varying $N_C$ is equivalent to changing the LIGO rate by the reciprocal factor.} 
\label{fig:constantK}
\end{figure}

 \subsubsection{Effects of the mass gaps on the BHMF}
The lower mass gap (LG), and the upper mass gap (UG) (see Figure \ref{fig:constantK}) affect the mass function evolution and produce features at various scales. First, the absence of BHs in the LG reduces the number of BHs of all scales. This is because no BH beyond the LG can merge with BHs in the LG to produce a more massive BH. The size of this reduction depends on the size of the LG, but is degenerate with the normalization of the IMF. As such, it is difficult to conclude anything about the LG, or even infer its existence, through this phenomenon. 

Due to the self-similar nature of the constant coagulation kernel, one might expect that the resulting BHMF to also be self similar. However, the gaps in the IMF spoils this self-similarity. The plots of Figure \ref{fig:constantK} display a break at $M\sim10 M_\odot$, which we call the lower break (LB). This break is caused because the BH formation channel where two BHs within the LG merge to form a BH beyond the LG is missing. Because the largest BH that could be formed by this channel is twice the largest BH in the LG, the LB is located at $M=2M_{\rm{max\; LG}}$, where the largest BH in the LG, $M_{\rm{max\; LG}}=5 M_\odot$. Changing $M_{\rm{max\; LG}}$ results in pushing the LG to larger masses. If detected, the existence of the LB can be used to diagnose both the existence and size of the LG.  

The interaction of the LG and the UG generates a break at $M\sim 60 M_\odot$ in Figure \ref{fig:constantK}. Because the most massive BHs in the IMF cannot merge with BHs in the LG, there is a dearth of BHs of mass $M_{\rm{min\; UG}}< M < (M_{\rm{min\; UG}} + M_{\rm{max\; LG}})$, where $M_{\rm{min\; UG}}$ is the most massive BHs in the IMF (the start of the UG). As the mass scale of the UB encodes the mass scale of the LG, an observation of the UB can be used to indirectly measure the size of LG.

\begin{figure}
 \centering
\includegraphics[width=3.3in]{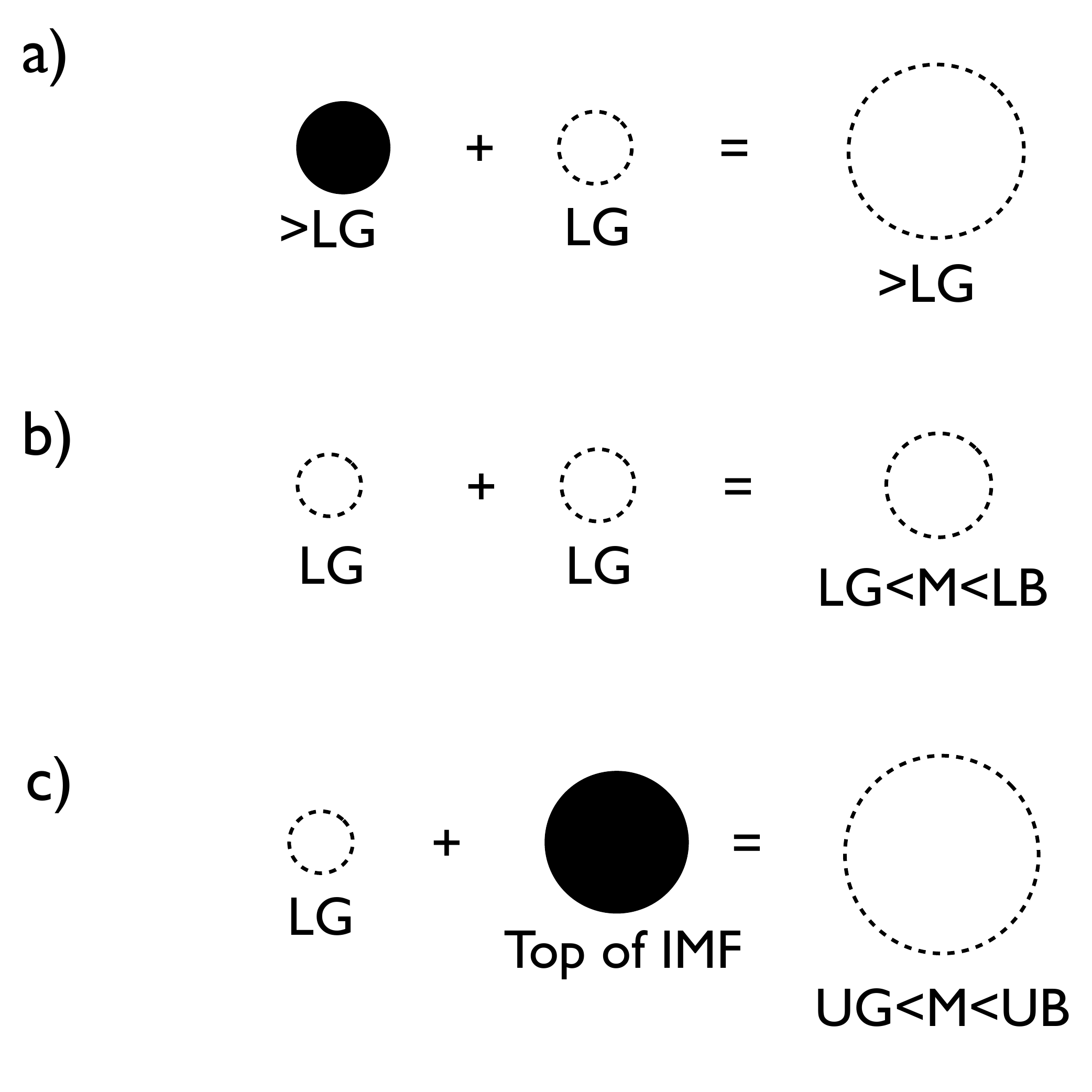} 
\caption{Missing BH formation channels due to the existence of the lower mass gap (LG) and the upper mass gap (UG). Case (a) shows that the number of BHs at all scales are lowered because no BH can merge with BHs in the LG to form a larger BH. Case (b) depicts the missing channel responsible for the break at $M=10 M_\odot$ in Figure \ref{fig:constantK}, which we call the lower break (LB). Because BHs generated by the mergers of two BHs within the LG is missing, there is a dearth of BH of mass $5 M_\odot  < M < 10 M_\odot$. Case (c) shows the missing channel that results from the interaction of LG and UG. Because BHs from the top of the IMF cannot merge with BHs within LG, there is a dearth of BHs with mass $50 M_\odot < M <60 M_\odot$, causing the break at $M=60 M_\odot$ in Figure \ref{fig:constantK}.} 
\label{fig:MissingChannels}
\end{figure}

The dearth of BHs that caused the LB and UB is also responsible to generating many more weaker breaks. Through a similar mechanism as was discussed in the previous paragraphs, anytime there is a dearth of BHs over a certain mass scale, there is a break due to there being fewer mergers than if the dearth is not present. However, these successive breaks are very weak, and are most probably not observable. Figure \ref{fig:MissingChannels} depicts all of the missing channels discussed in this section. 

 \subsubsection{Effects of varying the number of BHs per cluster}
\begin{figure}
\centering
\includegraphics[width=3.5in]{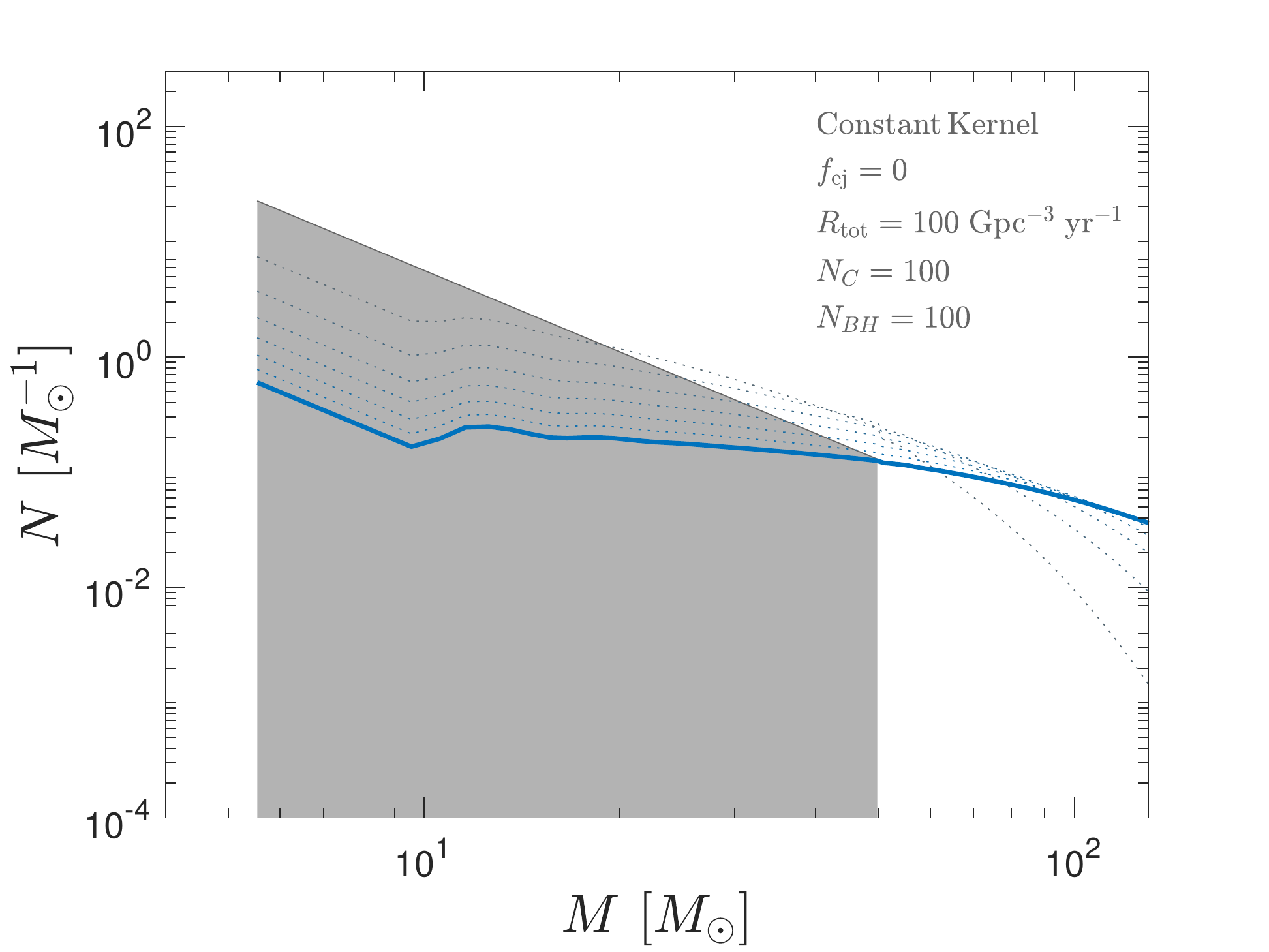}
\caption{The evolution of the BHMF starting from the IMF (black) to $10$ Gigayears (solid blue) for a cluster with $100$ BHs. Dotted blue lines represent the BHMF at intervening times. The LIGO rate is taken to be $100$ Gpc$^{-3}$ yr$^{-1}$, and the number of clusters per MWEG is taken to be $N_C = 100$. Lowering the number of BHs per cluster flattens the BHMF in a similar way as increasing the LIGO rate.} 
\label{fig:fewBH}
\end{figure}

Because the merger rate is calibrated to the observed LIGO rate, clusters containing fewer BHs need to have more efficient mergers than clusters containing more BHs. This is manifested in Equation (\ref{eq:RKrelationship}) as 
\beq
K \propto \frac{1}{N_{\rm BH}^2} \; .
\eeq
Due to this increase in efficiency, for the same LIGO rate, clusters can develop a flat BHMF if they contain few BHs. Figure \ref{fig:fewBH} shows the evolution of the BHMF over $10$ Gyr with the same $R_{\rm{LIGO}}$ as the first plot of Figure \ref{fig:constantK} with $N_{\rm{BH}}=100$. If one assumes that the BHMF is a power law, then situations as shown in Figure \ref{fig:fewBH} have to be excluded. Assuming a $R_{\rm{LIGO}}$ of $10^{-5}$ yr$^{-1}$ and $N_C=10$ requires each cluster to contain at minimum $\sim 1000$ BHs.
 
\subsection{Evolution with ejections} 
In the process of assembling a dynamical binary, or due to the merger kicks experienced by a merged BH, a star cluster is continuously losing BHs. We model the ejection of BHs from the system by introducing a source function, $S(M,t)$, that reduces the number of BHs of mass $M$ by a number that is proportional to the amount of mergers that produce BHs of mass $M$,
\begin{align}
&S(M,t) = \nonumber \\
&\;\; - \frac{f_{\rm ej}}{2} \int_0^M K(M-M',M') N(M-M',t)N(M',t) {\rm{d}}M' \; ,
\end{align}
%CHECK numbers and references
where $f_{\rm ej}$ is the ejected fraction. In effect, this source function parameterizes the phenomenon that for every merger, a fraction $f_{\rm ej}$ of the BHs are ejected. While we kept the parameter $f_{\rm ej}$ as a single number, in reality the recoil kicks of binary BHs depend on the spins of the individual BHs. For simplicity, we will neglect the spin dependence of $f_{\rm ej}$.

\begin{figure}
\centering
\includegraphics[width=3.5in]{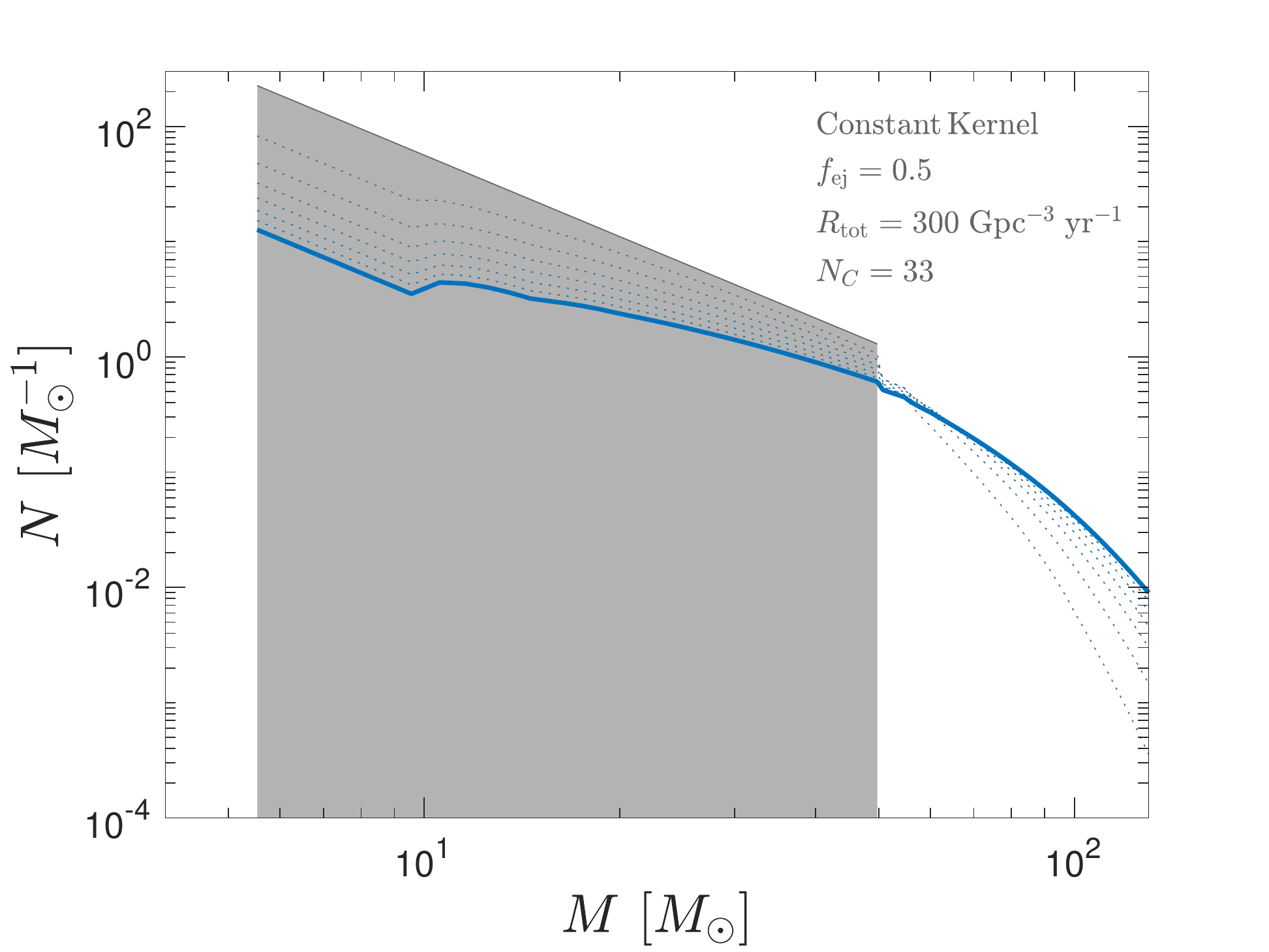}\\
\includegraphics[width=3.5in]{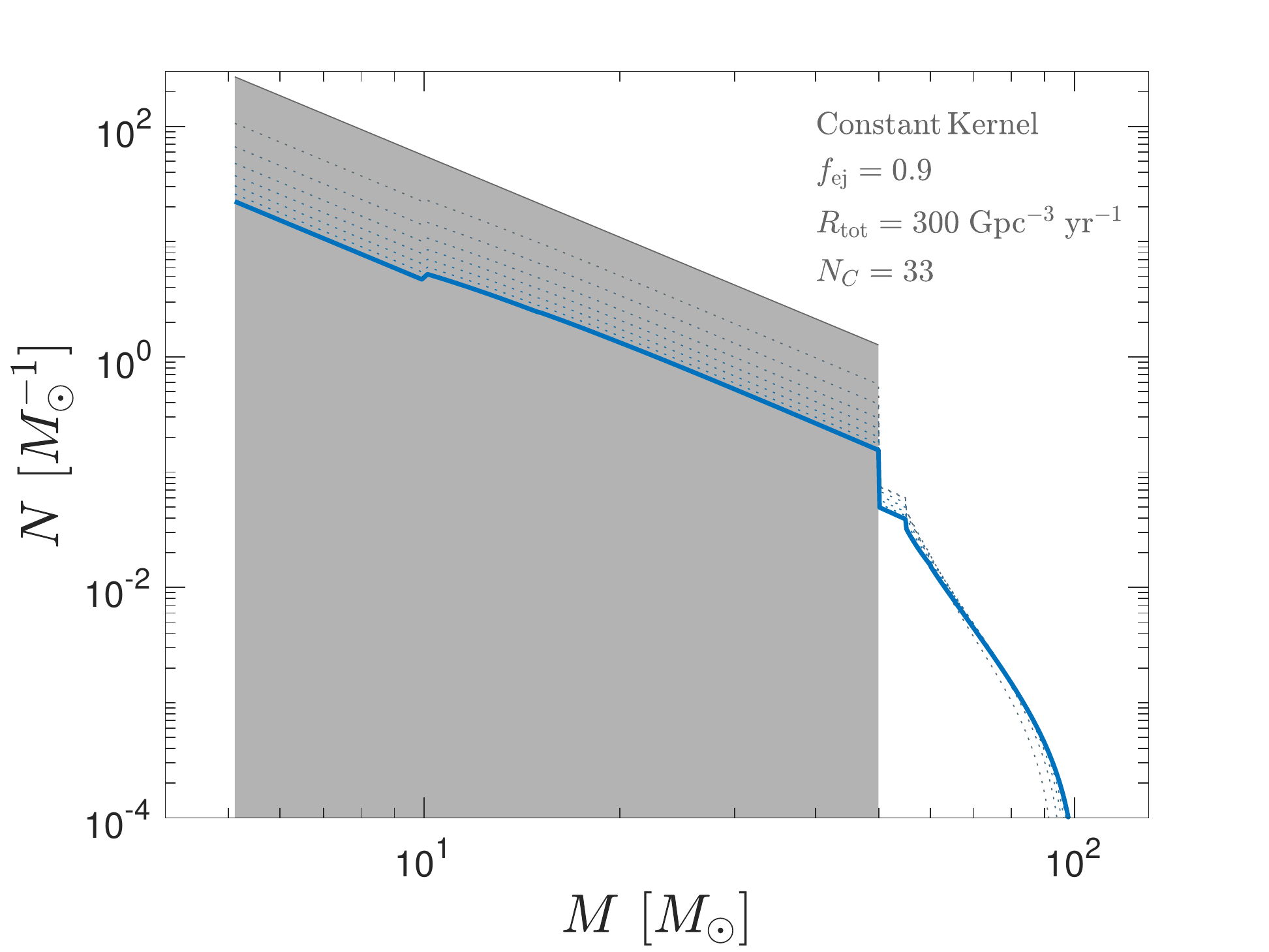}\\
\includegraphics[width=3.5in]{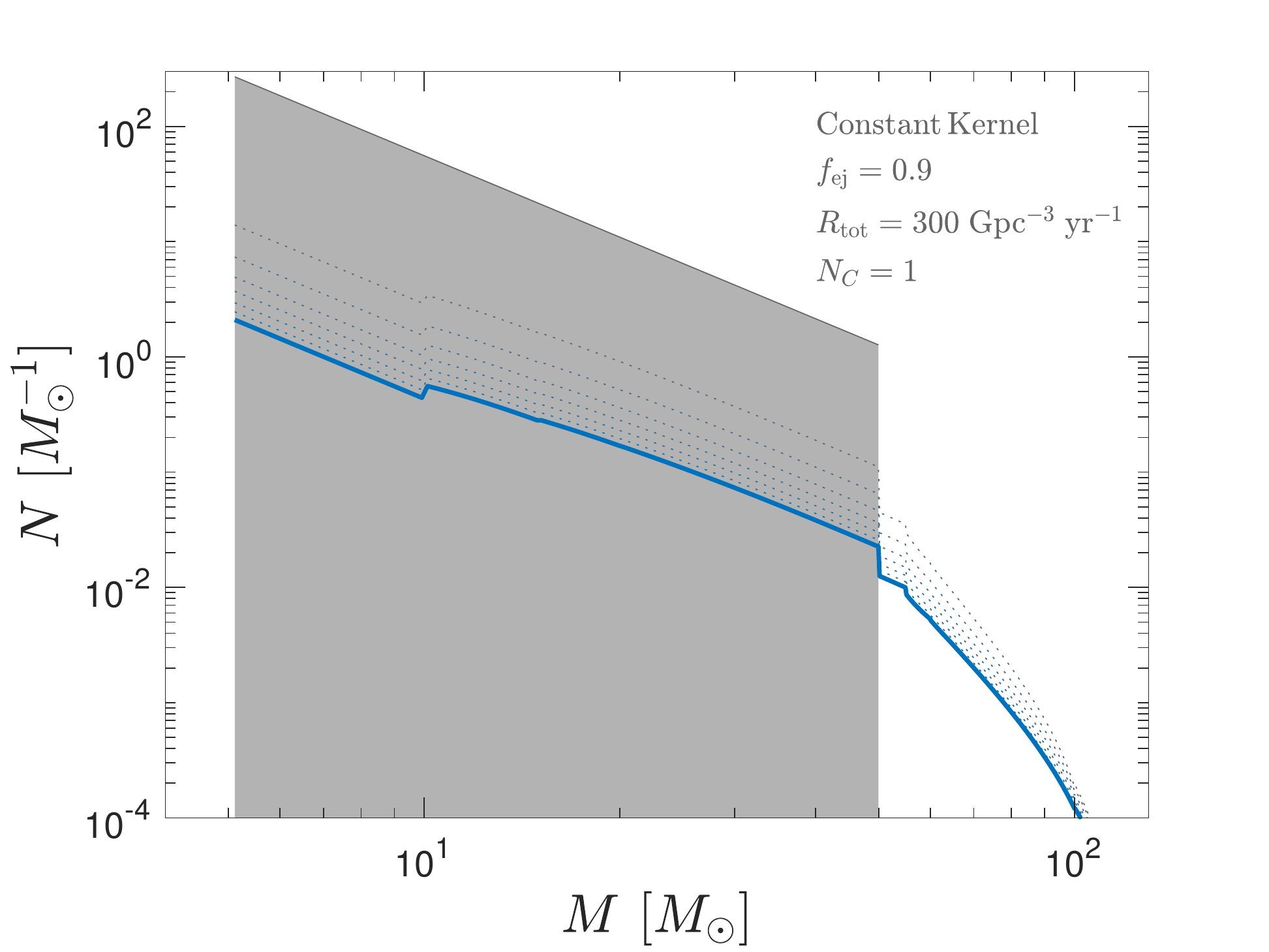}
\caption{The evolution of the BHMF starting from the IMF (black) to $10$ Gigayears (solid blue) for a cluster with $1000$ BHs. Dotted blue lines represent the BHMF at intervening times. The LIGO rate is taken to be $300$ Gpc$^{-3}$ yr$^{-1}$, and the number of clusters per MWEG is taken to be $N_C = 33$ (top, middle) and $N_C = 1$ (bottom). The ejection fraction is taken to be $f_{\rm ej} = 0.9$. Even for a LIGO rate per cluster of $300$ Gpc$^{-3}$ yr$^{-1}$, the BHMF fails to flatten in $10$ Gigayears.} 
\label{fig:eject}
\end{figure}

%Numerical simulations suggests that the ejected fraction can be very high [CITE Rasio:Rodriguez et al. (2015, 2016a), Chatterjee et al. (2016)], although the inclusion of Post Newtonian terms might curb the ejection fraction [CITE:CarlPN, Samsing].    
Note that this parameterization is agnostic towards the actual ejection mechanism. For a given merger, the two BHs that participate in the merger event can be kicked out during their assembly process, or the two BHs can merge, producing a gravitational wave recoil that ejects the merged BH from the cluster. Figure \ref{fig:eject} shows the evolution of the BHMF for a cluster with an ejection fraction of $f_{\rm ej}=0.5$ and $f_{\rm ej}=0.9$. 

There are a few main differences between a cluster without ejections and a cluster with efficient ejections. First, ejections lower the normalization of the BHMF, as there are less BHs at all scales. Next, ejections prevent the BHMF from being flattened. Indeed, as shown in the bottom plot of Figure \ref{fig:eject}, even a scenario with a merger rate at the top of the LIGO range ($R_{\rm tot}= 300$ Gpc$^{-3}$ yr$^{-1}$) fails to flatten the BHMF if $f_{\rm ej}$ is allowed to be very high. This allows clusters with low number of BHs ($N_{\rm BH} < 100$), or scenarios with very high merger rate per cluster to be consistent with the cutoff at $\sim 40 M_\odot$. 

In addition to the global properties described in the previous paragraph, efficient ejections also change the properties of the BHMF at certain scales. The LB turns into a step function when ejections are efficient, which might make its detection in the BHMF difficult. As seen in Figure \ref{fig:eject}, for the first few Gigayears, there is now a discontinuity at the end of the UB. Because channel (c) in Figure \ref{fig:MissingChannels} is missing, BHs with masses $50 M_\odot < M < 60 M_\odot$ (those between the start of the UG and the UB) are generally formed by fewer mergers than BHs generated beyond the UB. As for every merger there is a chance to be ejected out of the systems, BHs beyond the UB suffer more ejections than those below the UB. This discontinuity is a signature that the system is efficiently ejecting their BHs, and the drop is larger for higher $f_{\rm ej}$. However, the evolution of the coagulation equation tends to smooth out discontinuities, and the magnitude of the drop is heavily suppressed after $10$ Gyrs.

\section{Evolution with top-heavy kernels}

\begin{figure} [h!]
\centering
\includegraphics[width=3.5in]{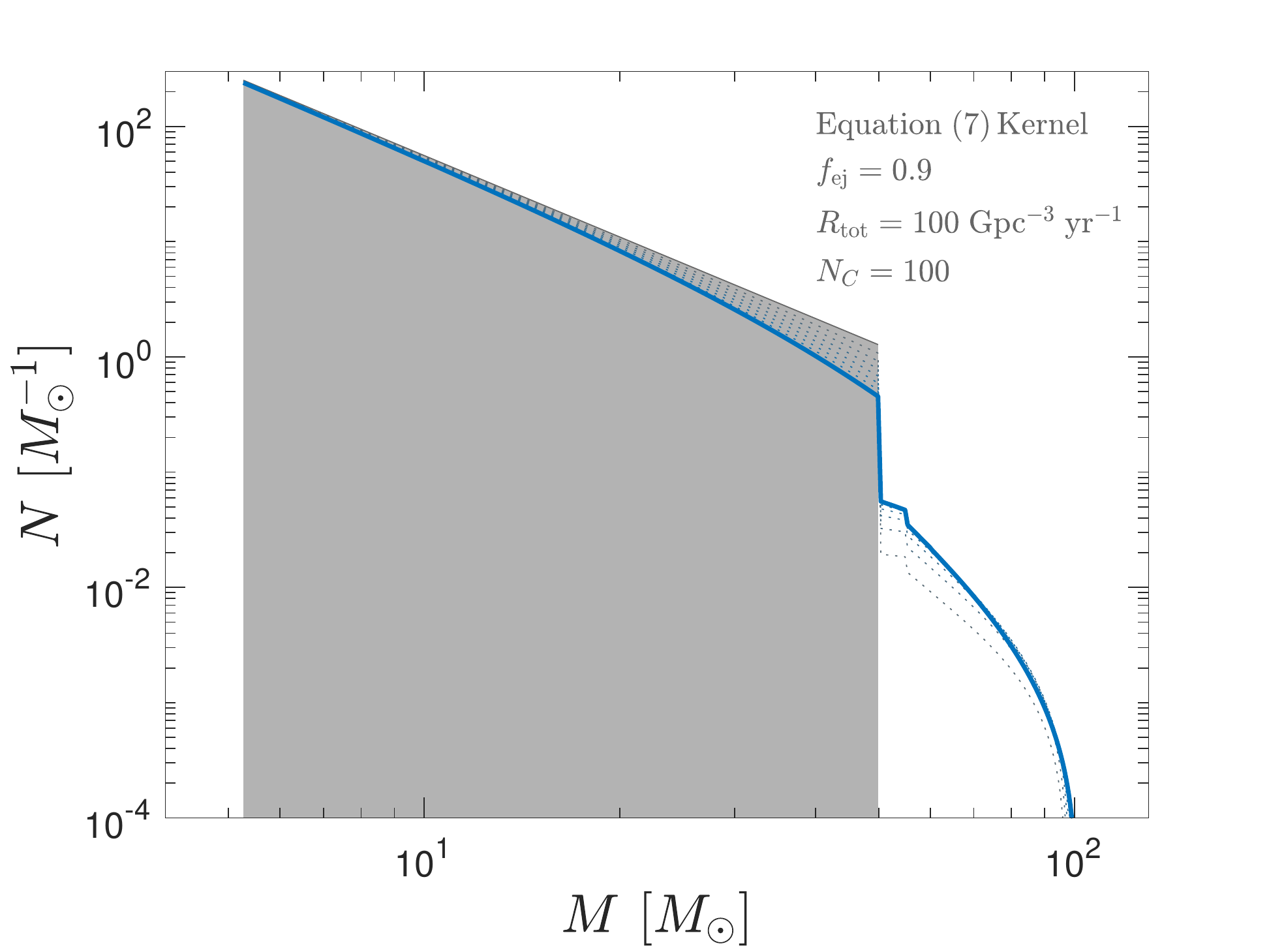}
\caption{The evolution of the BHMF starting from the IMF (black) to $10$ Gigayears (solid blue) for a cluster with $1000$ BHs for the top-heavy coagulation kernel given by equation (\ref{gravrad}). Dotted blue lines represent the BHMF at intervening times.. The LIGO rate is taken to be $100$ Gpc$^{-3}$ yr$^{-1}$, and the number of clusters per MWEG is taken to be $N_C \sim 100$. The ejection fraction is taken to be $f_{\rm ej} = 0.9$.} 
\label{fig:topheavy}
\end{figure}

Many phenomena responsible for dynamical mergers, such as gravitational capture, mass segregation, and 3-body relaxation are mass dependent. Therefore, we would expect that the coagulation kernel in equation (\ref{eq:Smol}) is in reality a function of mass, $K=K(M,M')$. While the actual form of the coagulation kernel depends on the dominant merging mechanism, it has to be symmetrical with respect to $M$ and $M'$. In general, this symmetry along with physical considerations forces the functional form of the coagulation kernel to be
\beq
K \propto (MM')^{\alpha} (M+M')^{\beta} \; ,
\eeq
with power-law indices $\alpha$ and $\beta$. The effectiveness of gravitational processes increases with increasing mass. Gravitational capture, for example, is more efficient for larger $M$ and $M'$. This implies that heavier BHs merge preferentially, and that $K$ is top-heavy. For example, the coagulation kernel due to gravitational radiation capture scales as \citep{2002ApJ...566L..17M}
\beq \label{gravrad}
K_{\rm{cap}} \propto (MM')^{15/14} (M+M')^{9/14}\; .
\eeq 
%While this lost of self-similarity is constrained in the low mass end by X-ray binary surveys (CITE:XrayBinary Survey),
Figure \ref{fig:topheavy} shows the evolution of the BHMF with the coagulation kernel given by equation (\ref{gravrad}). While the specific values of $\alpha$ and $\beta$ would matter for the numerical values of $f$, the salient features of the calculation is valid for general top-heavy kernels. The most important change introduced by the top-heavy kernel is the lost of the power-law behavior in the mass range $5 M_\odot < M < 50 M_\odot$. %As seen in the bottom plot of Figure \ref{fig:topheavy}, a very high LIGO rate creates a continuous BHMF that again, fails to flatten the BHMF.

\begin{figure} [h!]
\centering
\includegraphics[width=3.5in]{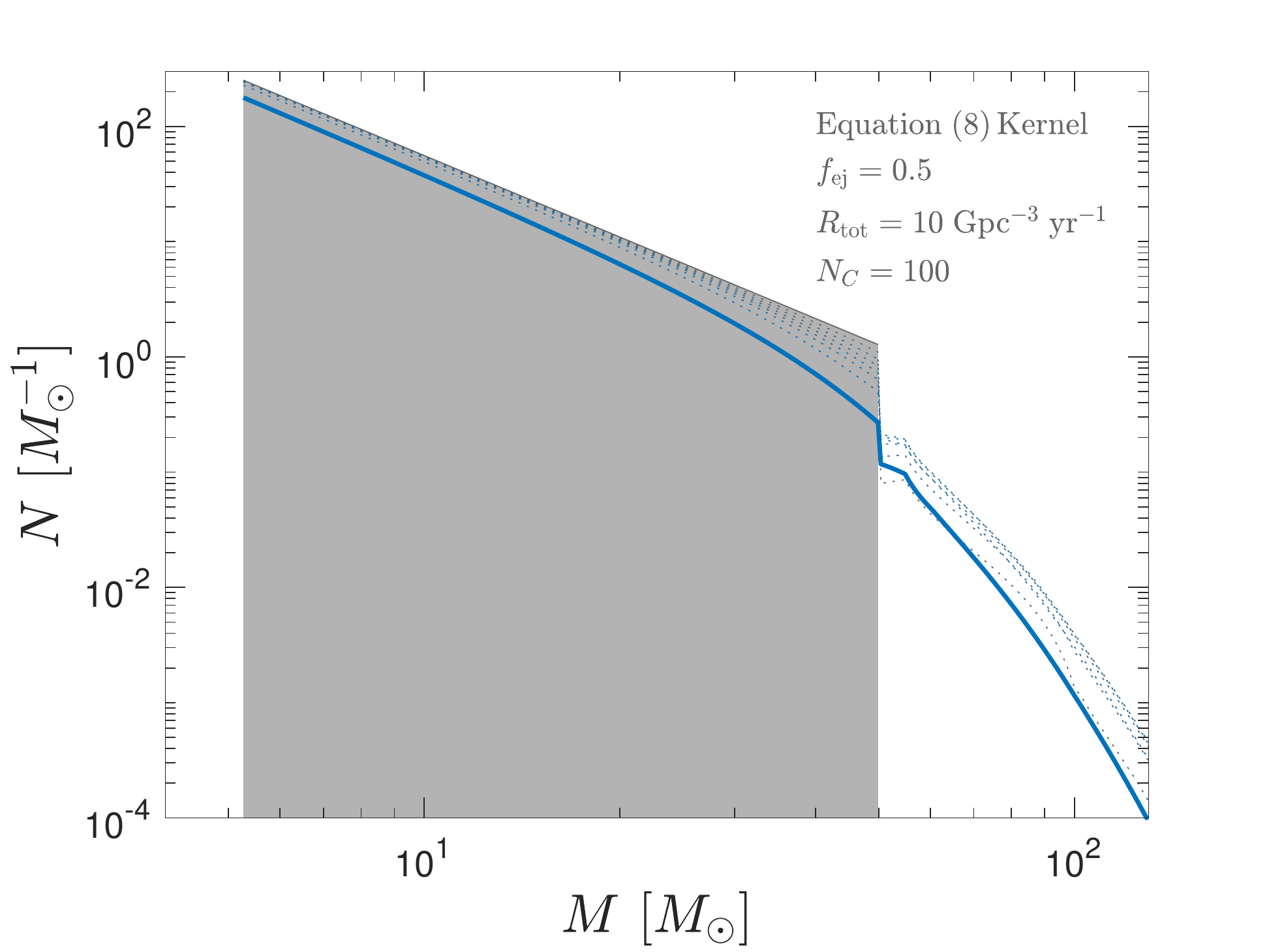}
\caption{The evolution of the BHMF starting from the IMF (black) to $10$ Gigayears (solid blue) for a cluster with $1000$ BHs for the 3-body coagulation kernel given by equation (\ref{eq:3body}). Dotted blue lines represent the BHMF at intervening times.. The LIGO rate is taken to be $10$ Gpc$^{-3}$ yr$^{-1}$, and the number of clusters per MWEG is taken to be $N_C \sim 100$. The ejection fraction is taken to be $f_{\rm ej} = 0.5$ and $M_{\rm max}=100 M_\odot$.} 
\label{fig:3body}
\end{figure}

Another example is the coagulation kernel from 3-body relaxation, which is computed through numerical simulations to scale as \citep{2016ApJ...824L..12O}
 \beq \label{eq:3body}
 K_{\rm{3-body}} \propto (M+M')^{4} \; .
 \eeq 
However, Ref. \cite{2016ApJ...824L..12O} did not fit for the $(MM')^\alpha$ component. In the coagulation equation, the $(MM')^\alpha$ term acts as a regularizer, and its absence generates a runaway growth of BHs that concentrates most of the cluster's mass in a single BH of extremely large mass $M \gtrsim 1000 M_\odot$. Observationally, we do not see such runaway growth. Thus, in order for this kernel to be consistent with observational bounds in the absence of the $(MM')^\alpha$ term, there must be some maximum $M_{\rm max}$ above which this kernel is suppressed. We impose this regularization by setting the kernel to be $(M_{\rm max} + M')^4$ for $M>M_{\rm max}$. While this introduces a new parameter to the problem, Figure \ref{fig:3body} shows that even a very conservative choice of $M_{\rm max}=100 M_\odot$, a significant population of BHs can be formed within the UG. Indeed, the use of this kernel does not change the main qualitative features of the other kernels, which is the possibility of intermediate mass BH seed formation in dynamical clusters. This echoes a previous result showing that in nuclear star clusters it is possible to obtain BHs in the intermediate mass ranges through multiple mergers \citep{2016ApJ...831..187A}. Our calculations extend this conclusion to the statement that globular clusters are also capable of producing intermediate mass BHs.

\section{Conclusion}
Through evolving a coagulation equation, we have shown that the BHMF in clusters could evolve to fill the gap in the IMF of BHs at $50 M_\odot < M < 130 M_\odot$. Further, we have found that the upper range of the LIGO rate is not consistent with the dearth of BHs with masses $M>40 M_\odot$ reported by \cite{2017ApJ...851L..25F} unless ejection is efficient. The coagulation equation also implies that the mass gap between the most massive neutron stars and the least massive BHs produces potentially observable features at larger scales. In addition, we show that that for top-heavy kernels, the mass function between $5 M_\odot < M < 50 M_\odot$ is driven away from self-similarity, and that a power-law will not be sufficient to fit the BHMF in this regime. With parameters consistent with realistic globular clusters, we showed that it is possible to form intermediate BH seeds through mergers of smaller BHs.  

%\todo{Very nicely written!! I think the main thing that needs to be done is annotate the figures (maybe by hand by a vector-graphics software that lets you add text into eps/pdf files). I think we need to label the regions UG, UG, etc. otherwise it is too hard quickly see what you are talking about. Also, I think we have to list the important parameters that describe the differences in the simulations in each figure: e.g. $N_C$, $f_{\rm ej}$, `constant kernel'... so someone can immediately see what are the intrinsic differences in all the plots without reading the captions. Otherwise at a quick glance they seem too repetitious/overwhelming.  }
%\todo{Need to also add units to figures, and y axis should be $N$ not $f$, right?}

\section*{Acknowledgements}
The authors would like to thank the anonymous referee, Chris Belczynski, and Tuguldur Sukhbold for comments on the manuscript. Support (PM) for this work was provided by NASA through Einstein Postdoctoral Fellowship grant number PF7-180164 awarded by the \textit{Chandra} X-ray Center, which is operated by the Smithsonian Astrophysical Observatory for NASA under contract NAS8-03060. This work was supported in part by Harvard's Black Hole Initiative, which is funded by a grant from the John Templeton Foundation.

\bibliography{BibFile_smol.bib}

\appendix

\section{Coagulation equation numerical solver}
\label{appendix1}

The coagulation equation is solved using a finite volume method based on 
\cite{2013arXiv1312.7240K}. The coagulation equation
can be written in conservative form as a function of $G(M,t)\equiv M\times N(M,t)$:
\begin{equation}
\partial_t G + \partial_M J(G) = MS
\end{equation}
where
\begin{equation}
J(G) = \int_0^M\int_{M-u}^{M_{\rm max}} \frac{K(u,v)}{v}G(t,u)G(t,v)\,dv\,du
\end{equation}
is the mass flux across mass bins. $G$ is a conserved quantity in the absence of source terms and conserved by our numerical finite-volume method.

The solution is discretized into mass bins and in time as $G^N_i(t_k)$
which represents the mean value of $G(t_k,M)$ in the mass bin $[M_i,M_{i+1})$ at time $t_k$.
The mass bin has center $M_{\rm{mid}(i)}$.
The left boundary flux is zero: $J_1^N(t_k)=0$.
In general, the flux $J_r^N(t_k)$ across each discrete boundary $x_r$
can be computed as follows, by considering the
aggregation of $M_{\rm{mid}(i)}$ and $M_{\rm{mid}(j)}$.
For each fixed $r$, and a fixed $i$ such that $M_{\rm{mid}(i)}<M_r$, then 
each $j$ such that $M_{\rm{mid}(j)} \geq M_r-M_{\rm{mid}(i)}$ 
gives a contribution to the flux of:
\begin{equation}
\Delta x G_i^N(t_k)\int_{M_j}^{M_{j+1}} \frac{K(M_{\rm{mid}(i)},y)}{y}G_j^N(t_k)\,dy
\end{equation}
A small exception occurs for the lowest $j$, where the lower limit of integration is
$M_{\rm{mid}(j)}$ instead of $M_j$.
The integral is evaluated for an arbitrary kernel numerically using a quadrature rule.

The equations are explicitly evolved from time step $t_k$ to $t_{k+1}$ as:
\begin{equation}
G^{N\prime}_i = G^N_i(t_k) + (\Delta t/2)\times M_iS_i(t_k,G^N_i(t_k))
\end{equation}
\begin{equation}
G^{N\prime\prime}_i = G^{N\prime}_i + \Delta t\frac{J^N_{i+1}-J^N_i(t_k)}{\Delta x}
\end{equation}
\begin{equation}
G^N_i(t_{k+1}) = G^{N\prime\prime}_i + (\Delta t/2)\times M_iS_i(t_k,G^{N\prime\prime}_i)
\end{equation}
that is, adding the source term in two half-steps which sandwich the flux term to result in a second-order method.

\end{document}